\documentclass[prl,aps,amssymb,nofootinbib,twocolumn,showpacs]{revtex4}
\usepackage{amsmath}
\usepackage{amssymb}
\usepackage{amsthm}
\usepackage{amsfonts}
\usepackage{algorithmic}
\usepackage{enumerate}
\usepackage{latexsym}
\usepackage[dvips]{graphicx}

\newcommand{\beq}{\begin{equation}}
\newcommand{\eneq}{\end{equation}}
\newcommand{\beqs}{\begin{equation*}}
\newcommand{\eneqs}{\end{equation*}}

\input{epsf}

\begin{document}

\tolerance 10000

\title{Coherent Quantum Tunneling of Bose-Einstein Condensates in Optical Lattices.}

\author { Zaira Nazario$^\dagger$ and
David I. Santiago$^{\dagger, \star}$ }

\affiliation{ $\dagger$ Department of Physics, Stanford
University,Stanford, California 94305 \\ 
$\star$ Gravity Probe B Relativity Mission, Stanford, California 94305}

\begin{abstract}
\begin{center}

\parbox{14cm}{ We study inter-well coherent quantum tunneling of Bose-Einstein
condensates (BEC's) in optical lattices. Irrespective of whether the BEC is a 
superfluid in the whole lattice, it may or may not be a superfluid within each 
well of the lattice depending on whether the ratio of the well kinetic energy 
to the interaction energy is small or large. When the bosons in each well are 
superfluid, we 
have Josephson tunneling among the wells driven by the phase difference between
the superfluids in the different wells. For typical lattice parameters the 
Josephson tunneling rate will be somewhat enhanced from the naive estimate of
(tunneling energy scale)$/\hbar$. Whether it is enhanced or not will depend
on lattice parameters, but it can never be zero. When the bosons in each well
are not superfluid, there is coherent quantum tunneling of essentially free
bosons between the wells. This tunneling is proportional to the boson 
number difference between the wells.  }

\end{center}
\end{abstract}


\date{\today}

\maketitle

In the present note we will study Bose-Einstein condensate (BEC) tunneling in 
an optical lattice. It has been verified experimentally in one dimensional 
optical lattices that there is coherent quantum tunneling in BEC's\cite{becj}. 
We will try to take an in-depth look at tunneling in optical lattices in 3 
dimensions. We take the well tunneling amplitude, or hopping, to be $t$. In 
order for the tunneling to be relevant, this energy scale needs to be larger 
than the kinetic energy scale for the whole lattice:

\beq
T_L = \frac{\hbar^2}{2 m \Omega^{2/3}}
\eneq

\noindent where $m$ is the mass of the bosons and $\Omega$ is the volume of 
the whole lattice. If the kinetic energy of the lattice overwhelms
the hopping, the tight-binding approximation with bosons hopping from lattice 
site to lattice site with tunneling $t$ is no longer valid. 

On dimensional grounds one expects the tunneling rate to be 
proportional to $t/\hbar$. On the other hand, the proportionality factor should
be a universal dimensionless function of the number of lattice sites $N_L$, the
number of bosons $N$ and/or the number of bosons in the condensate $N_0$, and 
the different energy scales in the problem. Besides $t$ and $\hbar^2 / 2 m 
\Omega^{2/3}$, there is the on-site repulsion at each well $U$, the well depth
$\varepsilon$, and the kinetic energy scale of each well:

\beq
T_w =\frac{\hbar^2}{2 m V^{2/3}}
\eneq

\noindent where $V$ is the volume of each well $\sim a^3$ with $a$ the 
interlattice spacing. We, of course, have $N_L V = \Omega$. Thus the tunneling 
rate or ``Josephson'' current is

\beq
J= \frac{t}{\hbar} F\left(N_0, N_L, \frac{t}{\varepsilon}, 
\frac{t}{U}, \frac{T_w}{U}\right)
\eneq

The universal function $F$ will behave differently in different physical 
regimes. For example, if the wells are very deep, $\varepsilon \gg t$, the 
tunneling rate should be exponentially suppressed $J \sim \exp(- \alpha 
\varepsilon/t)$ with $\alpha$ a positive numerical constant. Such a case is 
uninteresting for the collective lattice dynamics of the boson system and will 
not be considered here. Hence we will never write the dependence of the 
function $F$ on $t/\varepsilon$. We will study tunneling when there are a 
large number of bosons in the adjacent sites between which tunneling is 
occurring. This can correspond to the case  where there is a very large number 
of bosons in the whole lattice or when the condensate is placed at two sites 
and its tunneling behavior monitored.

The BEC on an optical lattice can be described by the Bose Hubbard Hamiltonian.
Since we have a large number of bosons per well, there will be a BEC in each 
well with its own Bose Hubbard Hamiltonian. There will be a tunneling term $t$ 
between adjacent wells. We then have 

\begin{align}
\nonumber \mathcal{H}&= \sum_{\vec k} \epsilon_{\vec k} a_{\vec k}^\dagger
a_{\vec k} + \frac{U}{N_L} \sum_{\vec k_1 \; \vec k_2  \vec q}
a_{\vec k_1 + \vec q}^\dagger a_{\vec k_2 - \vec q}^\dagger a_{\vec k_1} 
a_{\vec k_2} \\
\nonumber & +\sum_{\vec k} \epsilon_{\vec k} \bar{a}_{\vec k}^\dagger
\bar{a}_{\vec k} + \frac{U}{N_L} \sum_{\vec k_1 \; \vec k_2  \vec q}
\bar{a}_{\vec k_1 + \vec q}^\dagger \bar{a}_{\vec k_2 - \vec q}^\dagger 
\bar{a}_{\vec k_1} \bar{a}_{\vec k_2} \\
& - t \sum_{\vec k} \left( a_{\vec k}^\dagger \bar{a}_{\vec k} + 
\bar{a}_{\vec k}^\dagger a_{\vec k} \right) \label{bosonh}
\end{align}

\noindent where $a_{\vec k}^\dagger, \; a_{\vec k}$ are bosonic creation and
destruction operators with momentum $\vec k$ on the right well, while the 
``barred'' operators correspond to the left well. $\epsilon_{\vec k}$ is the
kinetic energy of a boson within each well. The wells are taken as large boxes.
This approximation is valid as wells in optical lattices are fairly large
($a \sim 100$'s of nm) in typical lattices\cite{ari1}.

As long as the bosons within each well are Bose condensed, the individual
well Hamiltonian can be diagonalized by the Bogolyubov method\cite{bog}.
If we keep the leading term in the interactions, we obtain the reduced
Bogolyubov Hamiltonian

\begin{align}
\nonumber \mathcal{H}&= \sum_{\vec k} \tilde{\epsilon}_{\vec k} 
a_{\vec k}^\dagger a_{\vec k} + \frac{U N_0}{N_L} \sum_{\vec k}
a_{\vec k}^\dagger a_{-\vec k}^\dagger a_{-\vec k} a_{\vec k} \\
\nonumber & +\sum_{\vec k} \tilde{\epsilon}_{\vec k} \bar{a}_{\vec k}^\dagger
\bar{a}_{\vec k} + \frac{U N_0}{N_L} \sum_{\vec k}
\bar{a}_{\vec k}^\dagger \bar{a}_{-\vec k}^\dagger 
\bar{a}_{-\vec k} \bar{a}_{\vec k} \\
& - t \sum_{\vec k} \left( a_{\vec k}^\dagger \bar{a}_{\vec k} +
\bar{a}_{\vec k}^\dagger a_{\vec k} \right) \label{redbog}
\end{align}

\noindent with $\tilde{\epsilon}_{\vec k}=\epsilon_{\vec k} + 2 U
N_0/N_L$. The number in the condensate, $N_0$, is determined uniquely
by the number of bosons in each well, $N$, and the repulsion $U$. We see
that the relevant parameter that sets the scale for the interaction
for the BEC is the combination $UN_0/N_L$. Hence the universal
dimensionless function that determines the tunneling current
enhancement does not depend on $U, N_0$ and $N_L$ independently, but
rather

\beq 
F\left( N_0,N_L,\frac{t}{U},\frac{T_w}{U} \right) = F\left(
\frac{tN_L}{UN_0}, \frac{T_wN_L}{UN_0} \right)
\eneq

The reduced Bogolyubov Hamiltonian (\ref{redbog}) is diagonalized by
the Bogolyubov canonical transformation

\beq
b_{\vec k}=u_{\vec k}a_{\vec k}+v_{\vec k}a_{-\vec k}^\dagger \qquad
b_{\vec k}^\dagger=u_{\vec k}a_{\vec k}^\dagger+v_{\vec k}a_{-\vec k}
\eneq

\beq
\bar{b}_{\vec k}=u_{\vec k}\bar{a}_{\vec k}+v_{\vec
k}e^{-i\theta}\bar{a}_{-\vec k}^\dagger \qquad \bar{b}_{\vec
k}^\dagger=u_{\vec k}\bar{a}_{\vec k}^\dagger+v_{\vec
k}e^{i\theta}\bar{a}_{-\vec k}
\eneq

\noindent where $\theta$ is the relative phase between the superfluid
order parameters, which is in general nonzero, and

\beq
u_{\vec k}^2=\frac{1}{2}\left( 1+\frac{\tilde{\epsilon}_{\vec
k}}{E_{\vec k}} \right) \qquad v_{\vec k}^2=\frac{1}{2}\left(
1-\frac{\tilde{\epsilon}_{\vec k}}{E_{\vec k}} \right)
\eneq

\noindent with

\beq
E_{\vec k} \equiv \sqrt{\tilde{\epsilon}_{\vec
k}^2-\frac{4U^2N_0^2}{N_L^2}}=\frac{\hbar
k}{\sqrt{2m}}\sqrt{\frac{\hbar^2 k^2}{2m} + \frac{4UN_0}{N_L}}
\eneq

\noindent being the dispersion of the quasiparticle-like excitations
of the system with $\epsilon_{\vec k}=\hbar^2 k^2/2m$. As $k
\rightarrow 0$ the dispersion becomes phononic, which leads to
dissipationless flow, i.e. superflow, by the Landau
argument\cite{landau}. We note that atomic BECs are superfluids as
they have a nonzero sound speed\cite{becs} and reduced scattering at
long wavelengths\cite{scat1, scat2, scat3}. The dispersion at large
$k$ becomes $\hbar^2 k^2/2m$, which leads to dissipative flow. Notice
that the crossover between dissipative and nondissipative flow
happens at $k \simeq \sqrt{8mUN_0/\hbar^2N_L}$. This defines the
coherence length of the superfluid $\xi \simeq
\sqrt{\hbar^2N_L/8mUN_0}$.

When the system is probed at length scales $L > \xi$ it responds
collectively as a superfluid, while at length scales $L < \xi$ it
behaves like a gas of independent bosonic particles, thus responding
dissipatively. We note that an infinite system would become superfluid
for any $U$ no matter how small, as one can always go to a long enough length 
scale to see the system respond collectively as a superfluid. In here we do not
have in mind infinite systems but finite wells of superfluid. In this
case of finite systems, if the kinetic energy scale of the well, $T_w$, is 
greater than the effective interaction, $UN_0/N_L$, the system {\it will not} 
be a superfluid as the linear size of the system, $V^{1/3}$, is less than
the coherence length of the superfluid.  Independent of whether the bosons are
superfluid in each well,  the system could still be a superfluid within the 
whole lattice.  

For the full lattice, when the kinetic energy scale is set by
tunneling between sites ($\epsilon_{\vec k}\sim tk^2a^2$), the
coherence length is $\xi \sim a\sqrt{N_Lt/4N_0U}$. The size of the
lattice is, of course, $L=N_La$ and the condition for superfluidity
($\xi < L$) is $U/t > 1/4N_0N_L$. We see that either for a very large
system or very large number of bosons, it becomes arbitrarily easy to
become superfluid in the sense that an arbitrarily small $U$ will
order the boson fluid into a superfluid at low enough temperature. Conversely,
as we make the system small, or we reduce the number of bosons, it becomes 
harder to order into a superfluid, which will show up experimentally as 
increased dissipation rates. 

Even though a high enough $U$ can order the system 
into a superfluid, too high a $U$ will make it into a Mott 
insulator\cite{fisher,us}. The transition to the Mott insulator can be 
continuous\cite{fisher} or discontinuous\cite{us}. The continuous transition 
occurs when there is a superfluid\cite{fisher} in each well and $U$ prevents 
Josephson tunneling between wells. The continuous transition only happens for 
a commensurate number of bosons per site. When the bosons in the wells are not 
superfluid {\it within} wells the transition is discontinuous and happens 
irrespective of whether the number of bosons per site is integer or 
not\cite{us}. The second transition always happens when the number of bosons 
per site or $U/t$ are suffciently small as in these cases $\xi >a$. 

We now proceed to evaluate the current operator\cite{joseph}

\begin{align}
\nonumber \hat{J} &\equiv \frac{d}{dT}(N-\bar{N}) =
-\frac{i}{\hbar}\left[ N-\bar{N},H_t \right] \\ &=
-\frac{2it}{\hbar}\sum_{\vec k}\left( a_{\vec k}^\dagger\bar{a}_{\vec
  k} - \bar{a}_{\vec k}^\dagger a_{\vec k} \right)
\end{align}

\noindent where $H_t=-t\sum_{\vec k}\left( a_{\vec
k}^\dagger\bar{a}_{\vec k} + \bar{a}_{\vec k}^\dagger a_{\vec
k}\right)$ is the tunneling part of the Hamiltonian and $T$ denotes
time. We note that the ground state of the reduced Hamiltonian
(\ref{redbog}) when $t=0$ is an outer product of the Bogolyubov ground
states for each well, which we denote as our ``vacuum'' state
$|0\rangle$. The excited states of such a system are denoted by
$|n\rangle$ and are constructed by applying any number of Bogolyubov
creation operators to the vacuum state. The energies of such states,
$E_n$, will be the sum of the energies of each of the operators. When
$t \neq 0$ there will be tunneling between the two wells, which can
be calculated accurately from linear response theory as long as the
number of atoms within the wells is large. For such a calculation the
only relevant excited states are two quasiparticle states, one in each
well, with opposite momenta, i.e. $|n\rangle=\bar{b}_{\vec k}^\dagger
b_{-\vec k}^\dagger |0\rangle$, with energies $E_n=2E_{\vec k}$. This
calculation was first done by B. D. Josephson for Cooper paired
fermionic superconductors\cite{joseph}. We will follow his calculation
closely. When the wells are superfluid, the tunneling current is
dominated by coherent tunneling of Bogolyubov correlated pairs.

In order to calculate in linear response theory\cite{pines}, we turn
on $H_t$ adiabatically from time $T=-\infty$, starting from the ground
state of the noninteracting system at such a time. If we expand the
wavefunction in terms of the noninteracting eigenstates
$|\Psi\rangle=\sum_n a_n(T)\exp{(-iE_nT/\hbar)}|n\rangle$, we obtain
the well known formula
$a_n(T)=\delta_{n,0}-i\exp{(iE_nT/\hbar)}\langle n|H_t|0\rangle/(iE_n
+ \epsilon)$, where the vacuum state energy $E_0$ has been chosen to be 0. 
The parameter $\epsilon$ was introduced to control the adiabatic
switching of the interaction. The causal limit $\epsilon \rightarrow
0^+$ should be always understood. Our interaction is turned on fully
at time $T=0$ and the wavefunction of the interacting system is given
by $|\Psi(0)\rangle = |0\rangle + |\delta\varphi\rangle$ where\cite{pines}

\beq
|\delta\varphi\rangle = -\sum_{n\neq 0}\frac{|n\rangle\langle
 n|H_t|0\rangle}{E_n - i\epsilon}
\eneq

The Josephson current is given by
$\langle\Psi(0)|\hat{J}|\Psi(0)\rangle=\langle0|\hat{J}|\delta\varphi\rangle+$
complex conjugate. With

\beq
|\delta\varphi\rangle=t\sum_{\vec k}\frac{u_{\vec k}v_{\vec
 k}}{2E_{\vec k}-i\epsilon} \left( e^{-i\theta}b_{\vec k}^\dagger
 \bar{b}_{-\vec k}^\dagger + b_{-\vec k}^\dagger \bar{b}_{\vec
 k}^\dagger\right)
\eneq

\noindent we obtain

\beq
J=-\frac{2t^2}{\hbar}\sin{\theta}\sum_{\vec k}\frac{u_{\vec
 k}^2v_{\vec k}^2}{E_{\vec k}}
\eneq

\noindent analogous to the fermionic superconducting
case\cite{joseph}. We note that if the phase difference between the
order parameters, $\theta$, is 0 there is no Josephson
tunneling. Since $u_{\vec k}v_{\vec k}=UN_0/E_{\vec k}N_L$, we obtain
for the enhancement factor $F=J\hbar/t$

\beq
F=-\frac{2tU^2N_0^2}{N_L^2}\sin{\theta}\sum_{\vec k}\frac{1}{E_{\vec k}^3}
\eneq

\noindent In order to estimate this quantity we approximate the
quasiparticle dispersion as phononic, $E_{\vec k}^2 \simeq
2UN_0\hbar^2 k^2/mN_L$, and we cut off the high momenta at $k_c$ of
the order of the inverse coherence length. Converting the sum into an
integral

\beq 
\sum_{\vec k}\frac{1}{E_{\vec k}^3} \simeq \frac{1}{(2\pi)^3}\left(
\frac{mN_LV^{2/3}}{2UN_0} \right)^{3/2} \int_{k < k_c}
\frac{d^3\vec{k}}{k^3}
\eneq

\noindent Note that this integral is infrared divergent but the
divergence gets cut-off by the inverse linear well size $V^{1/3}$. We
finally obtain for the enhancement

\beq\label{f}
F=\frac{1}{16\pi^2}\frac{tN_L}{UN_0}\left( \frac{UN_0}{T_wN_L}
\right)^{3/2}\; \ln{\frac{4UN_0}{T_wN_L}}
\eneq

In order for the system to be superfluid at all within a well we must
have $V^{1/3} > \xi$, otherwise {\it there will not be} Josephson
tunneling per se and the calculation of tunneling {\it will have} to
be done for noninteracting bosons trapped in wells. This condition
implies that $UN_0/T_wN_L > 1$. Therefore, $F >
tN_L/16\pi^2UN_0$. For typical lattices\cite{ari1}, $t$ is of the
order of around kHz's. $N_L$ is about 50, the number of bosons is about 
5000, so that the number of Bose condensed atoms per well $N_0 \sim
N$ and thus $N_0/N_L \sim 100$. Values of $U$ are more
uncertain, but they range from tens to a hundred  Hz, so that $t/U >
10$. We finally obtain $F > 2$. Therefore, when we have few hundred bosons 
per site, Josephson tunneling will be enhanced from the single 
particle tunneling value $t/\hbar$. Of course, for other values of the lattice
parameters, it need not be enhanced. In extreme cases, it can be
enhanced quite a bit. Since the logarithm in expression
(\ref{f}) can never be negative due to the precondition for Josephson
tunneling to be possible at all ($V^{1/3}> \xi$), $F$ is single valued.  We
then see that measurements of the Josephson ``enhancement'' $F$ can be used to 
estimate the values of the on-site repulsion $U$ if the other experimental
parameters are known.

We now study the case when the linear size of the wells is too small 
($V^{1/3}< \xi$). In this case the interaction cannot order the system into a 
superfluid, and its only effect is a renormalization of the mass of the 
bosons. Hence the system is described by the Hamiltonian (\ref{bosonh})
with $U=0$ and $m$ the renormalized mass of the bosons. There is no Josephson 
tunneling per se as the system is not superfluid. On the other hand, there can 
be coherent quantum tunneling of the (nonsuperfluid) boson liquid between 
wells. 

The Hamiltonian (\ref{bosonh}) with $U=0$ is diagonalized by the 
canonical operators

\beq
A_{\vec k} = \frac{a_{\vec k} + \bar{a}_{\vec k}}{\sqrt{2}} \qquad
\bar{A}_{\vec k} = \frac{a_{\vec k} - \bar{a}_{\vec k}}{\sqrt{2}}
\eneq

\noindent to yield

\beq
\mathcal{H}= \sum_{\vec k} \left \{(\epsilon_{\vec k} -t)A_{\vec k}^\dagger
A_{\vec k} +  (\epsilon_{\vec k} + t)\bar{A}_{\vec k}^\dagger \bar{A}_{\vec k} 
\right \} \label{diagh}
\eneq

\noindent The Heisenberg equations of motion lead to the following time 
dependent operators $A_{\vec k}(T) = A_{\vec k}\exp[-i(\epsilon_{\vec k} -t)T
/ \hbar]$ and $\bar{A}_{\vec k}(T) = \bar{A}_{\vec k}\exp[-i(\epsilon_{\vec k}
+t)T / \hbar]$. The common phase factor $-i\epsilon_{\vec k} T/ \hbar$ will be 
dropped as only phase differences affect the physics. From these time dependent
operators one immediately obtains the operators into which $a_{\vec k}$ 
$\bar{a}_{\vec k}$ and their conjugates develop under Hamiltonian evolution:

\begin{align}
\nonumber a_{\vec k}(T) = \cos\left( \frac{tT}{\hbar}\right)a_{\vec k} 
+ i \sin\left( \frac{tT}{\hbar}\right)\bar{a}_{\vec k} \\ \nonumber
\bar{a}_{\vec k}(T) = \cos\left( \frac{tT}{\hbar}\right)\bar{a}_{\vec k} 
+ i \sin\left( \frac{tT}{\hbar}\right)a_{\vec k} \\ \nonumber
a_{\vec k}^\dagger (T) = \cos\left( \frac{tT}{\hbar}\right)a_{\vec k}^\dagger 
- i \sin\left( \frac{tT}{\hbar}\right)\bar{a}_{\vec k}^\dagger \\
\bar{a}_{\vec k}^\dagger (T) = \cos\left( \frac{tT}{\hbar}\right)
\bar{a}_{\vec k}^\dagger -i \sin\left(\frac{tT}{\hbar}\right)a_{\vec k}^\dagger
\end{align}

\noindent The ``barred'' and ``unbarred'' operators are orthogonal raising and 
lowering operators as unitary Hamiltonian evolution is a canonical 
transformation thus preserving the commutation relations of the operators at 
time zero.

We can now calculate the coherent tunneling current between two wells.
We take the right well to have $(N + M)/2$ bosons and the left well with 
$(N - M)/2$ bosons. The initial wavefunction is then

\beq
| \Psi (0)\rangle = \Big [\frac{(a_0^\dagger)^{(N + M)/2} 
(\bar{a}_0^\dagger)^{(N - M)/2} }{(N/2+M/2)!(N/2-M/2)!} \Big ] |0\rangle
\eneq

\noindent where the subscript $0$ means the smallest momentum state into which 
the bosons condensed. The wavefunction at time $T$, $| \Psi (T)\rangle$, has 
the exact same dependence on the operators $a_{0}^\dagger (T)$, 
$\bar{a}_{0}^\dagger(T)$ that $| \Psi (0)\rangle$ has on 
$a_{0}^\dagger$, $\bar{a}_{0}^\dagger$. In order to obtain the 
coherent tunneling current $J$ we need to evaluate 
$\langle \Delta N(T) \rangle \equiv \langle \Psi (T)| \Delta N | 
\Psi (T)\rangle$ with $\Delta N \equiv a_{0}^\dagger a_{0} - 
\bar{a}_{0}^\dagger \bar{a}_{0}$ counting the excess number of bosons in the 
right well. Using

\begin{align} \nonumber
\Delta N =  \cos\left(\frac{2tT}{\hbar}\right)(a_{0}^\dagger(T) a_{0}(T) - 
\bar{a}_{0}^\dagger (T) \bar{a}_{0}(T)) \\ + i
\sin\left(\frac{2tT}{\hbar}\right) (\bar{a}_{0}^\dagger(T) a_{0}(T) - 
a_{0}^\dagger(T) \bar{a}_{0}(T))
\end{align}

\noindent we obtain the tunneling current 

\beq
J= \frac{d}{dT}\langle \Delta N(T) \rangle = \frac{2t}{\hbar}M \sin\left(
\frac{2tT}{\hbar}\right)
\eneq

\noindent where the periodicity is an artifact of having a two site system. 
The proportionality of the tunneling current on the initial number difference
of bosons between the sites, $M$, is not an artifact. Hence the tunneling 
current could be quite large if the number of bosons is very different between 
the different sites. The current can also be zero if the number of bosons is 
equal between wells. The fact that the current goes to zero for short times $T 
\rightarrow 0$ is also not an artifact since, for short times, the system 
cannot know if it is a closed two site system or a large system. 

In the present note we studied quantum tunneling of BEC's in optical lattices. 
For the case when bosons within each well are superfluid, we have Josephson 
tunneling among the wells  For typical lattice parameters the Josephson 
current will be somewhat enhanced from $t/\hbar$, but this need
not be so. The enhancement factor will depend on lattice parameters. When the 
bosons in each well are not superfluid, there is coherent quantum tunneling of 
essentially free bosons between the wells. This tunneling is proportional to
the boson number difference between the wells essentially due to Bose 
statistics.

\noindent {\bf Acknowledgements:} We thank Ari Tuchman and Mark
Kasevich for numerous and stimulating discussions. Zaira Nazario is a
Ford Foundation predoctoral fellow. She was supported by the Ford
Foundation and by the School of Humanities and Science at Stanford
University. David I. Santiago was supported by NASA Grant NAS 8-39225
to Gravity Probe B.

\end{document}